# Parametric Investigation Of Different Modulation Techniques On Free Space Optical Systems

Nauman Hameed [b], Tayyab Mehmood [a, 1], Anisa Qasim [a, 2]
[a] University College of Engineering & Technology, the Islamia University Bahawalpur, Pakistan.
[b] Department of Telecommunication Engineering, UET Taxila, Pakistan.
[a, 1] tayyabjatoibaloch@gmail.com, [b] nauman112te@gmail.com,
[a, 3] anisaqasim@gmail.com

*Abstract*—Free Space Optics systems (FSO) is one of the evolving wireless technologies. FSO is the only technology with highest data rates in wireless mode of operation but it suffers from bad weather conditions. In this work, analysis is carried out on FSO system having certain parameters constant using different modulation formats (i.e. RZ, NRZ, MDRZ, MODB and CSRZ). Impact of data rate, link range, input power and attenuation factor has been computed. Weather conditions are supposed to be nearly clear and suitable for FSO communication while taking attenuation factor up to 10dB/Km. Q-factor, received signal power and BER is calculated in all scenarios for obtaining an estimate of system performance. Results have shown that NRZ & RZ formats are in the lead until now with highest Q values.

*Keywords*— Free Space optics, Modulation formats, Q-factor, Received signal power.

## I. INTRODUCTION

Free Space Optics (FSO) also termed as Optical Wireless Technology is an emerging technology in which a light is used as a medium for communication. Although FSO is being used in satellite communication yet its significance can't be denied in terrestrial mode of operation. FSO was first used for military purposes due to its unmatched security, flexibility and ease of installation. Being operated at high carrier frequencies, FSO comes up with high data rates. Insensitivity to electro-magnetic interference and jamming, license free operation, commercial availability, high level data protection, high bit rates and low initial cost make FSO to be considered as a strong candidate in next generation broadband access networks [1]. FSO can be implemented where physical connection is not feasible. FSO has a very high impact of weather conditions (i.e. rain, dust, fog, haze, snow). Before its implementation in some specific area or region, average weather situation must be estimated. Different phenomenon affect light beam in open atmosphere such as turbulence, absorption, scattering and scintillation. Bad weather conditions result in poor transmission and eventually low BER and signal power [2]. FSO links with 7km length are in operation but it works best when distance is in hundreds of meters [3]. The FSO link with short distance works well because attenuation in some worst cases of weather reaches 80-120 dB/Km. High transmitter power is needed to cope with such high degree of attenuation. Optical amplifiers are also used to mitigate the effect of attenuation. FSO can be a good choice for access networks which span hundreds of meters [4]. FSO is allowing end users to access optical connectivity reliably and at low per user cost. In wireless mode of operation, FSO is the only technology with such high bandwidth. FSO has fascinating applications in different scenarios such as last mile access, connecting nearby buildings, short term mobile links, satellite communication, disaster recovery, temporary setup for events and many more. FSO is also being used with RF & Microwave links in hybrid [5]. Although having much advantages and innate qualities to be chosen among next generation networks, FSO still needs much improvement to survive in bad weather conditions. So, one thing that matters is transmitter configuration and modulation techniques used. Different modulation formats give dissimilar results and performance under different parameters. We have done a study and simulation based on certain parameters using various modulation techniques (i.e. RZ, NRZ, MDRZ, MODB, CSRZ). Experiments have shown that external modulation gives better performance and BER as compared to direct modulation [6].

## II. SIMULATION SETUP

The proposed FSO system's performance is studied based on different parameters (i.e. Data rate, range, transmit power). System simulation is performed in Optisystem and performance is analyzed using BER analyzer. It gives us Q-factor, eye diagram analysis and received power of transmitted signal at receiver end. The reference system block diagram is given in Fig. 1.

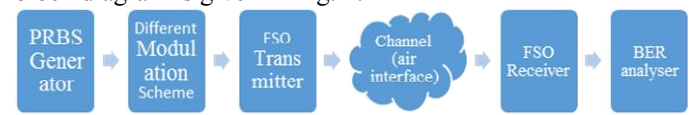

Fig. 1 Block diagram of reference FSO system

In figure. 1, different blocks are shown. Each block gives the idea behind its function being performed. First block visualizes as PRBS (Pseudo Random Bit Sequence) generator which creates data bits in binary form as 1's and 0's representing two different states generally known as On & Off. Pulse generators for various modulation formats (i.e. NRZ, RZ, MODB, MDRZ and CSRZ) are symbolized by second block. The output from this block is fed into DML (Directly modulated laser) which is the part of FSO transmitter as shown in third block. Channel air interface block between FSO transmitter and receiver is a medium of transmission. FSO receiver block contains APD (Avalanche Photodiode) receiver and a low pass Bessel filter of order 4 and 100dB depth. At the end, BER analyzer is present for computing Q-factor, eye diagram and received signal power with their respective graphs.

## III. REFERENCE SYSTEM CHARACTERISTICS:

| Design Parameters | Values |
|---|---|
| Operating Frequency | 1550nm |
| Modulator | Machzender Modulator |
| Sequence Length | 128 bits |
| Samples/bit | 64 |
| Number of Samples | 8192 |
| EDFA Gain | 10dB |
| Beam Divergence | 2mrad |
| Optical Detector | Avalanche Photodiode |
| Filter Type | Low Pass Bessel filter |
| Filter Order | 4 |

## IV. RESULTS AND ANALYSIS

The reference system is analyzed with different parameters under consideration to demonstrate their effects on overall system performance. Modulation formats selected for this design are NRZ, RZ, MODB, MDRZ and CSRZ. Parameters shown above in table 1 remain identical over all the simulations performed while various parameters changed (i.e. Data rate, link range, transmit power, Input aperture, Rx aperture, attenuation factor). All the effects of these parameters on system performance will be discussed in sections below.

## V. EFFECT OF DATA RATE

High data rate demand with low cost is increasing day by day. FSO is only wireless technology with huge data rates approaching certain Gbps. FSO works in a low range at very high data rates due to high impact of bad weather conditions. In access networks, FSO comes up as viable and secure solution.

In this scenario, data rate is altered to observe its effect on system performance and Q-factor. Data rate values range from 1Gbps up to 10Gbps. Link range selected is 2Km and attenuation factor is set at 10dB/Km. Graph in fig. 4 shows that at data rate ranging from 1 to 5Gbps, NRZ remains at top among all formats with highest Q-factor but after further increment in data rate, RZ format takes a lead. When data rate approaches to 6Gbps, the q-factor for MODB scheme reaches zero. At 7Gbps, q-factor for MDRZ format becomes zero as can be seen in the fig. 2. At peak data rate of 10Gbps, Q-factors are given as 3.57, 5.82, 0, 0 and 3.99 for NRZ, RZ, MODB, MDRZ and CSRZ respectively.

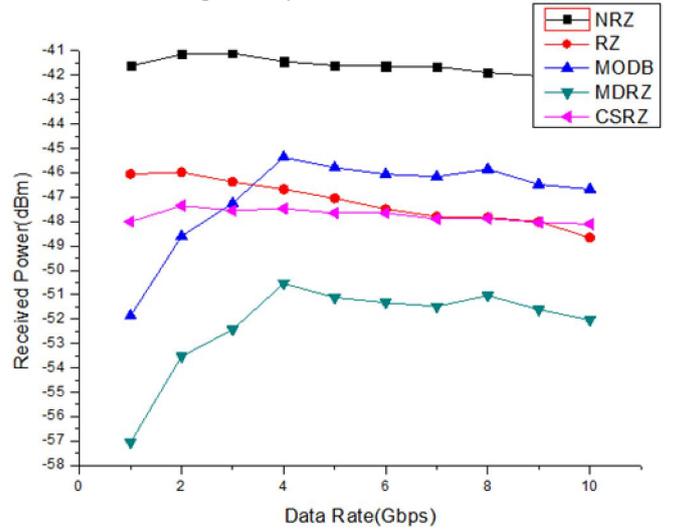

Fig. 3. Received Power vs. Data Rate using different modulation formats.

In fig.3, graph is plotted between data rate and power of signal at receiver. Analysis carried out for different modulation formats show that NRZ format gives maximum received power at all values of data rate ranging from 1 to 10 Gbps. A fair competition can be seen between RZ and CSRZ.

## VI. EFFECT OF LINK RANGE

As FSO systems are operated in open atmosphere therefore different factors affect the quality of FSO link. Link range is also one of the most important parameters to be considered before designing the FSO link. FSO links are present in practical applications operating in hundreds of meters of range. Links in range up to 7km are in operation but for long range links weather conditions must be evaluated before practical implementation of the link. In this proposed work, link range is altered from 1 to 10Km by keeping certain parameters constant. Input power is 10dBm with attenuation factor 2dB/Km. Light beam divergence angle is 3mrad with Input and Rx aperture sizes being 10 & 150cm respectively. Low pass Bessel filter and APD photodiode is used at the receiver.

Figure. 4 shows the Q-factor versus link range with different modulation formats. At link range up to 6Km, RZ format remains at the top with highest Q-factor among all formats. After that NRZ and RZ yields nearly identical Q values up to 10 Km of link range. Besides RZ and NRZ, MODB format is at third position. Q values are [21,11,6], [20,10,5], [14,7,4], [9,4,2] and [14,8,4] for transmission distance 8 Km, 9Km and 10Km in case of NRZ, RZ, MODB, MDRZ and CSRZ respectively. Q-factor shows that for long range links NRZ is best choice but for small ranges RZ performs well.

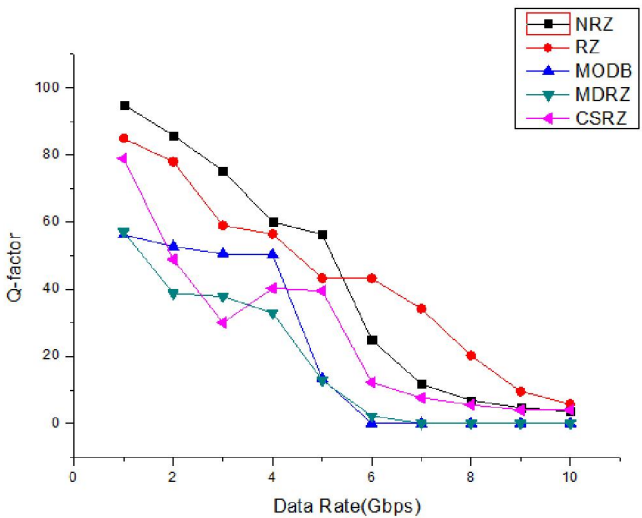

Fig. 2. Q-factor vs. Data rate using different modulation formats.

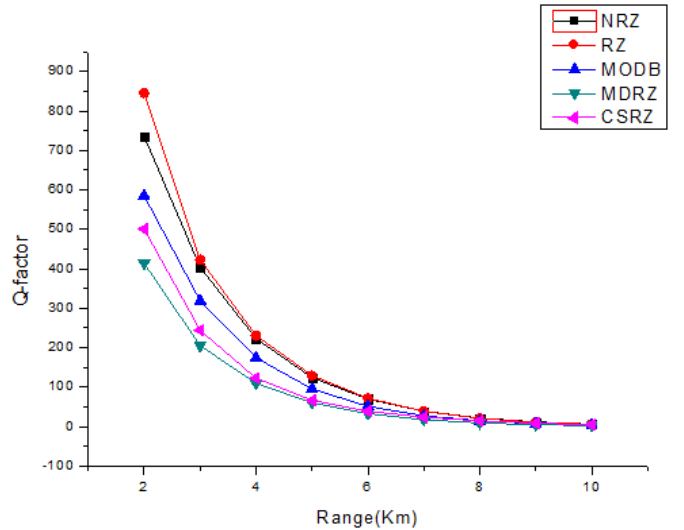

Fig. 4. Q-factor vs. Range using different modulation formats.

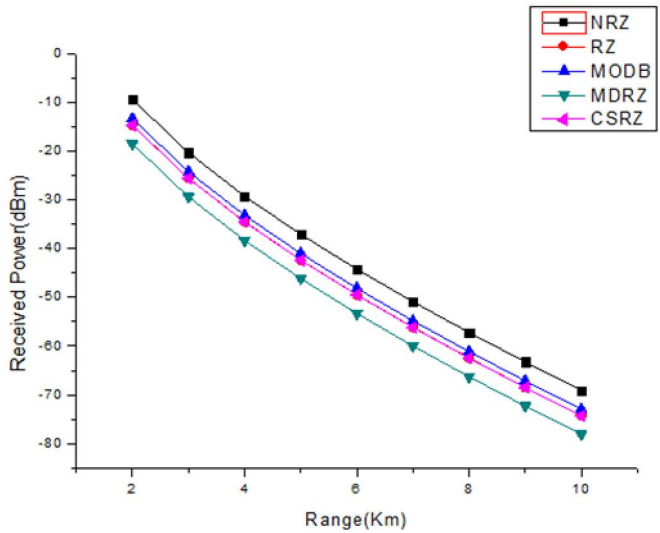

Fig. 5. Received Power vs. Range using different modulation formats.

Graph between received powers versus link range is shown in fig. 5. It shows that NRZ resulted in maximum received signal power while MODB comes at second position. RZ in this case results in a little bit low signal powers as compared to NRZ & MODB but identical to CSRZ format. BER values are [$1.194e^{-102}$, $5.89e^{-095}$, $1.524e^{-048}$, $2.75e^{-020}$, $8.59e^{-048}$], [$1.41e^{-30}$, $6.011e^{-28}$, $6.71e^{-015}$, $8.74e^{-07}$, $2.18e^{-017}$] and [$8.77e^{-010}$, $4.20e^{-09}$, $2.48e^{-05}$, $6.14e^{-03}$, $1.97e^{-06}$] for transmission distance 8 Km, 9Km and 10Km in case of NRZ, RZ, MODB, MDRZ and CSRZ respectively.

## VII. EFFECT OF INPUT POWER

Transmitter power is an important factor which plays vital role in communication system. Although we need to consume less power but a little bit power penalty can be good for reliability of the link. In this instance, we have applied various values of power at transmitter from 3dBm to 5dBm. Data rate, link range and attenuation are 10Gbps, 2Km and 10dB/Km respectively.

Figure. 6 clearly shows the evaluated results in which NRZ format is having highest Q values at all values of input power. After that RZ is also performing well. But for MODB and CSRZ, it can be seen that at values of power from -3 to 2dBm CSRZ is out performing the MODB format and after further increase in power up to 5dBm MODB takes the lead.

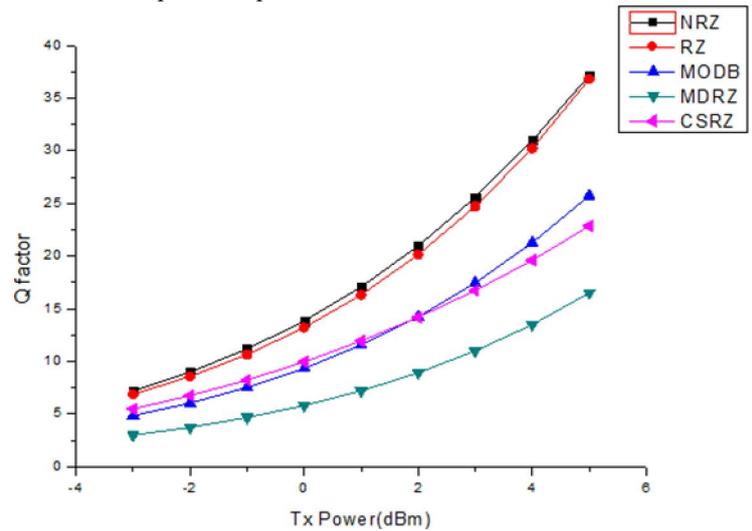

Fig. 6. Q-factor vs. Tx Power using different modulation formats.

Performance of MDRZ is satisfactory in this case. Q-factor values are [25, 30, 37], [24, 30, 36], [17, 21, 25], [11, 13, 16] and [16, 19, 22] for transmitter power 3dBm, 4dBm and 5dBm in case of NRZ, RZ, MODB, MDRZ and CSRZ respectively.

Figure. 7 shows the graph plotted between received signal powers versus transmitter power. With increase in transmitter power, there is an increase in received signal power. In this event, NRZ remains at the first position with peak received signal power at all values of Input power. MODB's performance is better in this case while RZ & CSRZ are having identical values. Received power values are [-55, -53, -51], [-60, -58, -56], [-59, -57, -55], [-64, -62, -60] and [-60, -58, -56] for input power 3dBm, 4dBm and 5dBm in case of NRZ, RZ, MODB, MDRZ and CSRZ respectively. All values are in dBm.

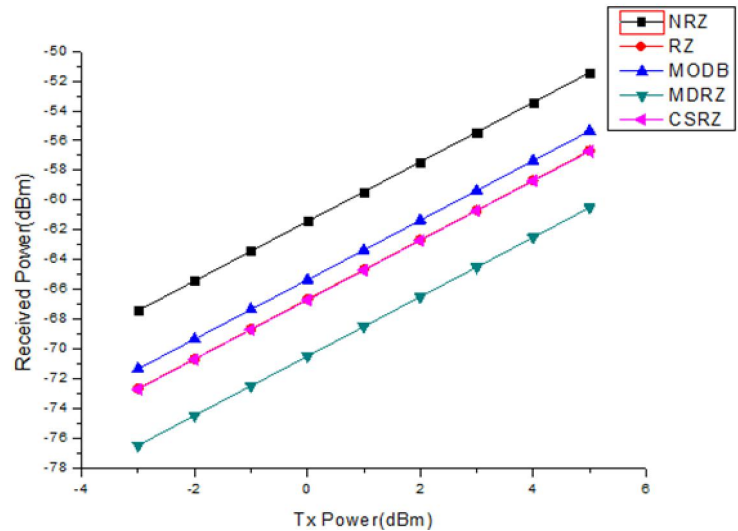

Fig. 7. Received Power vs. Tx Power using different modulation formats.

## VIII. EFFECT OF ATTENUATION FACTOR

Attenuation is one of the main factor in any communication system which needs to be compensated for reliable quality communication. In FSO system, it is having a very large effect according to diverse weather conditions. Operating frequency is set at 1550nm due to its less exposure to atmospheric attenuation [6]. Input power is 10dBm & beam divergence in this case is 3mrad. EDFA with 10dB gain is used. Transmission of data rate is at 10Gbps & link range is 3Km. At receiver side, APD photo-detector is used with 3dB gain, responsivity 1 A/W, ionization ratio 0.9 & dark current 10nA. Thermal noise is also considered. Six attenuation values (i.e. 1, 2, 3, 5, 7, 10dB/Km) are taken into account for investigation.

As figure. 8 shows the effect of attenuation on different modulation formats versus Q-factor. It can be seen that RZ format performs best under all values of attenuation. At low values of attenuation, there is a large difference in Q-factor values and it remains minor at large attenuation. There is a close competition between NRZ and RZ format. At attenuation value of 10dB/Km, Q-factor values for NRZ & RZ are 6.52 and 6.23 respectively. Min. BER values for NRZ & RZ are $3.48 \times 10^{-11}$ and $2.72 \times 10^{-10}$ respectively.

Figure. 9 shows the graph of received signal power versus attenuation. In this case, NRZ out performs all other modulation formats with highest signal power at receiver side. RZ and CSRZ formats are having almost identical received signal power values. MDRZ scheme gives lowest received signal powers at all values of attenuation.

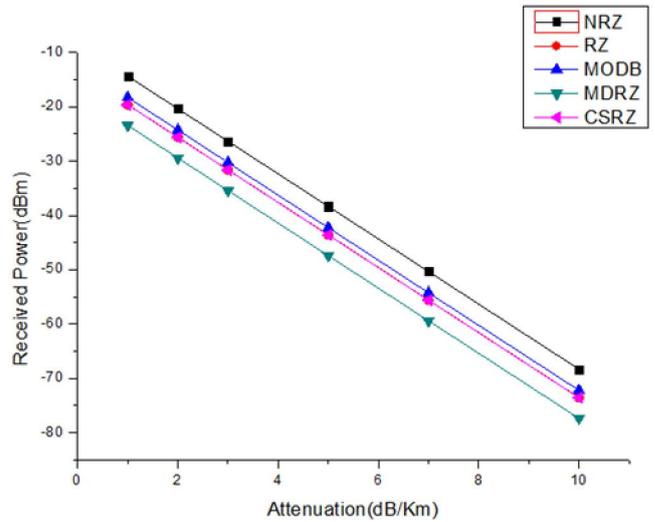

Fig. 9. Received Power vs. Tx Power using different modulation formats.

## IX. CONCLUSION

This work presents the complete analysis with certain parameters fixed while some parameters being changed in different cases. For low data rates, NRZ is a best choice but for higher data rates from 6 to 10 Gbps RZ remains at the top. For short link ranges, RZ seems to be a good candidate. It is concluded after observing various effects on the proposed system that NRZ & RZ formats are suitable for FSO systems according to their performance.

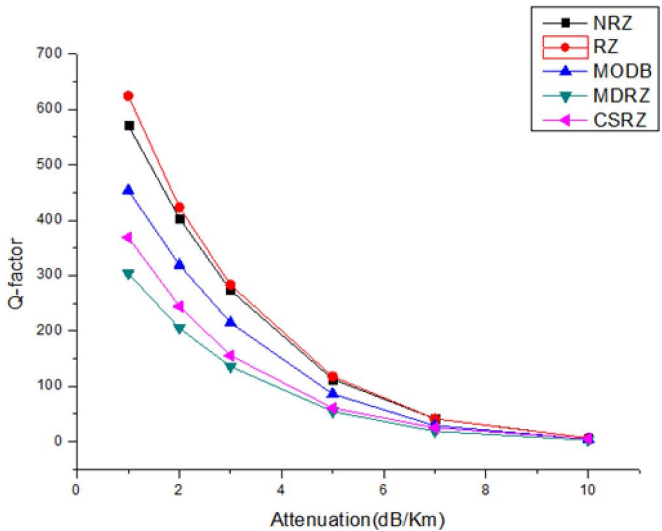

Fig. 8. Q-factor vs. Attenuation using different modulation formats.